\documentclass[reprint,a4paper,10pt,prl,aps,fleqn,superscriptaddress]{revtex4-1}

\usepackage[utf8]{inputenc}

\usepackage[tbtags]{amsmath}
\usepackage{amssymb}

\usepackage[load=accepted]{siunitx}
\usepackage{graphicx}

\usepackage{bm}

\newcommand{\lrangle}[1]{\ensuremath{ \big\langle #1 \big\rangle}}
\newcommand{\baryY}[2][]{\ensuremath{\lrangle{y{#1}}_{{#2}}}}
\newcommand{\baryYE}[2][]{\baryY[#1]{|E_{{#2}}|^2}}
\newcommand{\baryYET}[1][]{\baryY[#1]{\EDT}}
\newcommand{\EDT}{|\vec E_T|^2}
\newcommand{\EDE}[1]{|E_{{#1}}|^2}

\newcommand{\shifty}{\ensuremath{\Delta_y}}
\newcommand{\shiftT}{\ensuremath{\Delta_T}}

\newcommand{\OT}[1]{\mathcal{O}(\theta_0^{#1})}

\newcommand{\IINT}{\iint\limits}

\DeclareMathOperator{\dif}{d}

\renewcommand{\figurename}{Figure}
\newcommand{\figref}[1]{\figurename~\ref{fig:#1}}
\newcommand{\EQref}[1]{equation \eqref{#1}}
\newcommand{\EQsref}[1]{equations \eqref{#1}}

\newcommand{\CHANGE}[1]{#1}

\newcommand{\subrm}[1]{\ensuremath{_{\text{#1}}}}
\renewcommand{\vec}[1]{\ensuremath{\boldsymbol{#1}}}
\newcommand{\vect}[1]{\vec{\tilde {#1}}}
\newcommand{\vech}[1]{\vec{\hat {#1}}}

\newcommand{\br}{\linebreak[1]} %
\newcommand{\brOC}{\\} %

\begin{document}
\newlength{\figwidth}
\setlength{\figwidth}{\columnwidth}
\title{Observation of the geometric spin Hall effect of light}
\begin{abstract}
The spin Hall effect of light (SHEL) 
is the photonic analogue of the spin Hall effect occurring for charge carriers
in solid-state systems.
{This intriguing phenomenon
manifests itself when a light beam refracts at an air-glass interface
(conventional SHEL)},
or when it is projected onto an oblique plane,
the latter effect being known as geometric SHEL.
It amounts to a polarization-dependent displacement
perpendicular to the plane of incidence.
Here, we experimentally demonstrate 
the geometric SHEL for a light beam
transmitted across an oblique polarizer.
We find that the
spatial intensity distribution of the transmitted beam depends on
the incident state of polarization and its centroid
undergoes a positional displacement exceeding one wavelength.
This novel phenomenon is virtually independent from the material
properties of the polarizer and, thus, reveals universal features
of spin-orbit coupling.
\end{abstract}

\newcommand{\afMPL}{\affiliation{Max Planck Institute for the Science of Light, Erlangen, Germany}}
\newcommand{\afUni}{\affiliation{Institute for Optics, Information and Photonics, University Erlangen-Nuremberg, Germany}}
\author{Jan Korger}
\author{Andrea Aiello}
\email{andrea.aiello@mpl.mpg.de}
\author{Vanessa Chille}
\author{Peter Banzer}
\author{Christoffer Wittmann}
\afMPL
\afUni
\author{Norbert Lindlein}
\afUni
\author{Christoph Marquardt}
\afMPL
\afUni
\author{Gerd Leuchs}
\afMPL
\afUni

\date{\today}
\maketitle

Already in 1943, Goos and H\"anchen
observed that the position of a light beam totally reflected
from a glass-air interface differs from metallic reflection
\cite{Goos1947}.
This is the most well-known example of a longitudinal beam shift
occurring at the interface between two optical  media.
In honour of their seminal work, any such 
deviation from geometrical optics occurring in the
plane of incidence, is referred to as a Goos-H\"anchen shift.
Conversely, a similar shift occurring in a direction perpendicular
to the plane of incidence is known as Imbert-Fedorov shift
\cite{Fedorov1955,Imbert1972}.
These phenomena generally depend both on properties of the incident light beam
and the physical properties of the interface \cite{Aiello2012}.
Goos-H\"anchen and Imbert-Fedorov shifts have been observed at
dielectric \cite{Pillon2004},
semi-conductor \cite{Menard2009} and
metal \cite{Merano2007} interfaces.

\begin{figure}
 \includegraphics[width=\figwidth]{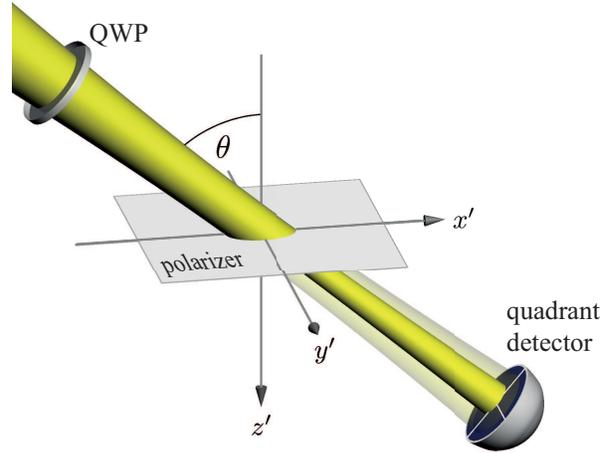}
 \caption{
Pictorial representation of
the
geometric SHEL. After the quarter wave-plate
(QWP) the beam is circularly polarized and passes
through a polarizer tilted by
an
angle $\theta$. The center of the
intensity distribution
of the transmitted beam appears shifted with respect to the
axis of the incident beam. Such a shift
can be
measured by a quadrant
detector put behind the polarizer.
The reference frame $\{\vec x', \vec y', \vec z'\}$ is aligned with the
polarizer surface.
    }
 \label{fig:beam3D}
\end{figure}

The Imbert-Fedorov shift
\cite{Fedorov1955,Schilling1965,Imbert1972} is an example of the
spin-Hall effect of light (SHEL),
the photonic analogue of spin Hall effects occurring for charge carriers in
solid-state systems \cite{Onoda2004,Hosten2008,Yin2013}.
SHEL is a consequence of the
spin-orbit interaction (SOI) of light, namely the coupling between
the spin and the trajectory of the optical field \cite{Dooghin1992,Leary2009,Liberman1992,Bliokh2004,Onoda2004,Bliokh2006,
Hosten2008,Aiello2008,Bliokh2008,Aiello2009a,Baranova1994,Bliokh2008b,Banzer2013}.
All electromagnetic SOI phenomena in vacuum and in locally isotropic 
media can be interpreted in terms of the geometric Berry phase and angular momentum 
dynamics \cite{Bliokh2012book}.
For a freely propagating paraxial beam of light, SOI effects vanish unless
a relevant breaking of symmetry occurs.
A typical example of such a symmetry break is the interaction of the beam with
an oblique surface
as in ordinary light refraction processes
(\figref{beam3D}).

Although the resulting phenomena essentially depend on the type of the
interaction with the surface \cite{Bekshaev2011},
there are some common characteristics that reveal universality in SOI of light due to the
geometry and the dynamical angular momentum aspects of the problem. Amongst the 
various observable effects resulting from the beam-surface interaction, 
the so-called \emph{geometric Hall effect of light} is virtually 
independent from the  properties of the surface 
\cite{Aiello2009a,Korger2011,Bekshaev2009,Bliokh2012b} and, therefore, 
represents the ideal candidate for studying above-mentioned universal features.
It amounts to a shift of the centroid of the intensity distribution represented by
Poynting-vector flow of the beam across the oblique surface of a tilted detector.
Direct observation of this effect as originally proposed depends critically on
the detector's response to the light field.
However, the question whether the response function of a real detector is indeed
proportional to the Poynting vector density is subject to a long-standing debate
\cite{Berry2009,Durnin1981,Braat2007}.

In this work, we implement an alternative scheme
\cite{Korger2011},
in which the occurrence of the beam shift is independent of the
detector response.
To this end, we send a circularly polarized beam of light across a 
tilted polarizing interface to demonstrate a novel kind of 
geometric Hall effect, in which the centroid of the resulting linearly
polarized transmitted beam undergoes a spin-induced transverse shift
up to several wavelengths. This is a spatial shift, which is independent of the distance the beam
propagated after interaction with the polarizer.
Our approach is different from the early proposal \cite{Aiello2009a} in
that it is observable with standard optical detectors.
Furthermore, it is different from the SHEL occurring in a beam passing
through an air-glass interface \cite{Hosten2008} since this geometric SHEL is
practically independent of Snell's law and the Fresnel formulas for the interface. 

\begin{figure}
    \includegraphics[width=\figwidth]{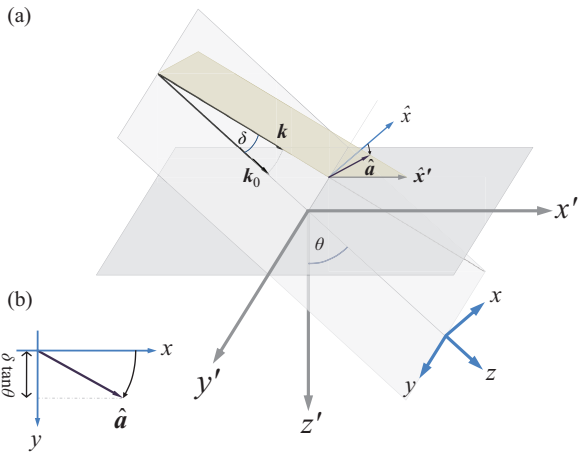}
	\caption{The geometric SHEL at a polarizing interface.
	(a) The horizontal 
	dark grey plane represents the polarizing interface with the Cartesian 
	reference frame $\{\vec x', \vec y'\equiv\vec y,\vec z' \}$ attached to it.
	 The axis $\vech x'$ 
	is taken parallel to the absorbing axis of the  polarizing 
	interface. $\theta \in [0,\pi/2)$ denotes the angle of incidence. The 
	central wave vector  $\bm{k}_0$ of the incident beam defines the 
	direction of the axis $\vech z$ of the frame  $\{\vec x,\vec y,\vec z \}$ attached to the 
	beam.
	It is instructive to study a wave vector $\vec k$ in the $z$-$y$-plane,
	rotated by an angle 
	$\delta \ll 1$ with respect to $\bm{k}_0$.
	(b) The unit vector $\hat{\bm{a}} \perp 
	\bm{k}$, representing the direction of the absorbed component of the incident field, lies
	in the common plane of $\bm{k}$ and $\vech{{x}}'$ (coloured light brown in the figure)
	and can be obtained, in the first-order approximation, from the
	rotation by the angle $\delta \tan \theta$ around $\bm{k}$, of the unit
	vector $\vech{{x}}$.}
    \label{fig:vectors}
\end{figure}

As for conventional SHEL, the physical origin of this geometric version 
resides in the SOI of light. In fact, we can describe this effect in terms of the geometric phase generated by the spin-orbit interaction as follows:
Think of the {monochromatic} beam as a superposition of many interfering plane waves
{with the same wavelength and different directions of propagation}.
When the beam passes through the polarizing interface, the plane wave components with different {orientations of their} wave vectors acquire different geometric phases determined by distinct local 
projections yielding to effective ``rotations'' of the polarization vector around the wave vector.
The interference of these modified plane waves produces a redistribution of the intensity spatial profile of the beam resulting in a spin-dependent transverse shift of the intensity centroid.

This effect can be better understood with the help of \figref{vectors} 
that illustrates the geometry of the problem. The incident 
monochromatic beam is made of many plane wave components with wave vectors $\bm{k}$ spreading around the central one  $\bm{k}_0 = k \vech{{z}}$,
which represents the main direction of propagation of the beam, where $|\bm{k}_0|= k = |\bm{k}|$. 
For a well-collimated  beam,
the angle 
$\delta$
between an arbitrary wave vector $\bm{k}$ and the central one $\bm{k}_0$ is, by definition, small:
$\delta \ll 1$. 
In the first-order approximation with respect to $\delta$, we consider
$\bm{k} = k (\vech{{z}} \cos \delta - \vech{{y}} \sin \delta) \cong k(\vech{{z}} + \kappa_y \, \vech{{y}})$,
where $\kappa_y \equiv k_y/k = -\sin \delta \cong -\delta$.
{
This wave vector is not the most general one, but,
due to the transverse nature of the phenomenon, it is sufficient to restrict the discussion to wave vectors lying in the
$yz$ plane.}
The either left- ($\sigma=+1$) or right-handed ($\sigma = -1$) circular
polarization of the incident beam is determined by the unit vector
$\vech{{u}}_\sigma = (\vech{{x}} + i \sigma \vech{{y}})/\sqrt{2}$
globally defined with respect to the axis $\vech z$ of the beam frame
$\{\vech x, \vech y, \vech z\}$. 
However, from Maxwell's equations it follows that the divergence of the electric field of a light
wave in vacuum must vanish. This requires that the polarization
vector
$\vech{{e}}_\sigma(\bm{k})$
attached to
each plane wave of wave vector $\bm{k}$ must necessarily be transverse,
namely $\bm{k} \cdot \vech{{e}}_\sigma(\bm{k}) = 0$.
This requirement is clearly not satisfied by $\vech{{u}}_\sigma$ for which one has $\bm{k} \cdot \vech{{u}}_\sigma/k \cong i \sigma \kappa_y/\sqrt{2} \neq 0$. 
Anyhow, the
correct expression for the polarization vector can be easily found
by subtracting from $\vech{{u}}_\sigma$ its longitudinal component: 
$\vech{{u}}_\sigma \rightarrow \vech{{e}}_\sigma(\bm{k}) \propto  \vech{{u}}_\sigma -  \bm{k} \left( \bm{k} \cdot \vech{{u}}_\sigma\right)/k^2 \cong
\vech{{u}}_\sigma -(i \sigma  \, \kappa_y/\sqrt{2})\vech{{z}}$. Now that we have properly modeled the polarization of the incident field, let us see how it changes when the beam crosses the polarizing interface.
 
A linear polarizer is an optical device that absorbs radiation polarized parallel to a given direction, say $\vech x'$, and transmits radiation polarized  perpendicular to that direction. 
The electric field of each plane wave component of the beam sent through
the polarizing interface, changes according to the projection rule
$\vech{e}_\sigma(\bm{k}) \rightarrow {\vech{e}_\sigma(\bm{k}) - 
\vech{{a}}(\vech{{a}}\cdot \vech{e}_\sigma(\bm{k}))} =  \vech{{t}}(\vech{{t}}\cdot \vech{e}_\sigma(\bm{k}))$,
where
$\bm{a} =  \vech{{x}}' - \vec{k} ( \vec k  \cdot \vech{x}')/k^2$
is the effective absorbing axis
with 
$\vech{{a}} = \bm{a}/| \bm{a}| \cong \vech{{x}} - \vech{{y}} \, \kappa_y \tan \theta$, and 
$\vech{t} = \vech a\times\vec k/k$
is the effective transmitting axis.
The amplitude of the transmitted plane wave is
$\vech{{t}}\cdot \vech{{e}}_\sigma(\bm{k}) \propto (1 - i \sigma \kappa_y \tan \theta)/\sqrt{2} \cong  \exp (-i \sigma \kappa_y \tan \theta)/\sqrt{2} $, where an irrelevant $\kappa_y$-independent overall phase factor has been omitted.
Therefore, as a result of the transmission, the amplitude of each plane 
wave component is reduced by a factor $1/\sqrt{2}$ and multiplied by 
the \emph{geometric phase} term $\exp (-i \sigma \kappa_y \tan 
\theta)$. Here, the ``$ \tan \theta$'' behavior of the phase, 
characteristic of the geometric SHEL \cite{Aiello2009a}, is in striking 
contrast to the typical ``$ \cot \theta$'' angular dependence of the 
conventional SHEL as, e.~g., the Imbert-Fedorov shift \cite{Liu2008}.  
However, the spin-orbit interaction  term $\sigma \kappa_y$, expressing the coupling  between the spin
$\sigma$
and the transverse momentum $\kappa_y$
of the beam, is characteristic of  both phenomena, thus revealing their common physical origin.
 
According to the Fourier transform shift theorem \cite{Goodman},
a linear phase shift in the wave vector domain introduces a translation in
the space domain.
Using this theorem, we can show that the
geometric phase term
$\exp (-i \sigma \kappa_y\tan \theta)$
leads to a beam shift along the $\vech y$-direction. 
This can be explicitly demonstrated by writing the incident circularly polarized paraxial 
beam as $\bm{\Psi}^\text{in}(y) = \vech{{u}}_\sigma 
\psi_\sigma^\text{in}(y)$, where only the dependence on
the relevant transverse coordinate $y$ has been displayed.
With $\widetilde{\psi}_\sigma^\text{in}(\kappa_y)$
we denote the Fourier transform of $\psi_\sigma^\text{in}(y)$.
The two functions are connected by the simple
relation
$\psi_\sigma^\text{in}(y) = \int \widetilde{\psi}_\sigma^\text{in}(\kappa_y)\exp(i k  \kappa_y y ) \text{d} \kappa_y $.
After transmission across the polarizing interface, the Fourier transform
$\widetilde{\psi}_\sigma^\text{in}(\kappa_y)$ of $\psi_\sigma^\text{in}(y)$ changes to
\linebreak[2]
$\widetilde{\psi}_\sigma^\text{in}(\kappa_y)\br\exp (-i \sigma \kappa_y \tan \theta)\br/{\sqrt{2}}$
\linebreak[2]
and the output field can be written as
\begin{align}
\nonumber
\psi_\sigma^\text{out}(y)
&= 
\frac1{\sqrt{2}}
\int \widetilde{\psi}_\sigma^\text{in}(\kappa_y)
\exp [i  k \kappa_y (y-\sigma\tan \theta/k)]
\text{d} \kappa_y
\\&=
\frac1{\sqrt{2}}
\psi_\sigma^\text{in}(y - \sigma \tan \theta /k)
\text.
\end{align}
This last expression clearly shows that the scalar amplitude of the output field is equal, apart from the $1/\sqrt{2}$ factor, to the amplitude of the input field transversally shifted by $\sigma\tan \theta/k$.
Finally, the position of the centroid of the transmitted beam is given by 
\begin{align}\label{eq:baryYideal}
\langle y \rangle = \frac{\int y \, |\psi_\sigma^\text{out}(y)|^2 \text{d} y}{\int |\psi_\sigma^\text{out}(y)|^2 \text{d} y}  =\frac{\sigma}{k} \tan \theta\text,
\end{align}
where $k = \frac{2\pi}{\lambda}$.

Eq.\ \eqref{eq:baryYideal} as derived above is valid for an {ideal} polarizer
with a high extinction ratio.
In this case, the description of the polarizer as a projection onto the effective transmission axis
$\vech t$ is adequate. At normal incidence ($\theta= 0$) such polarizers are readily
available.
However, working with tilted polarizing interfaces, we have to account for the fact
that the 
efficiency of the polarizer diminishes with larger angles $\theta$.
As a result of such loss of efficiency, the ``$\tan\theta$''-dependence of the
shift is modified as illustrated in Fig.\ \ref{fig:IF}.
This is explained in detail in the Supplemental Material,
where we introduce a phenomenological model for our real-world polarizer
\cite{Fainman1984,Korger2013_polarizer,Haus1993,Aiello2009d,Hung83,Hsu2004}.

\par
In the experiment, we use a Corning
Polarcor polarizer,
made of two layers of elongated and oriented silver nano-particles
embedded in a
$\SI{25}{\milli\metre}\times\SI{25}{\milli\metre}\times\SI{.5}{\milli\metre}$
glass substrate.
Directed absorption from these particles
effectively polarizes the transmitted
beam.
In order to avoid parasitic effects from the glass surfaces
($n_G = 1.517$), we have submerged the
polarizer in a tank with index matching liquid (Cargille laser liquid 5610,
$n_L = 1.521$).
Without this liquid, the effective tilting angle inside the glass polarizer would
be limited by Snell's law to $\arcsin(1/n_G)\approx
41\degree$.

\begin{figure}
 \includegraphics[width=\figwidth]{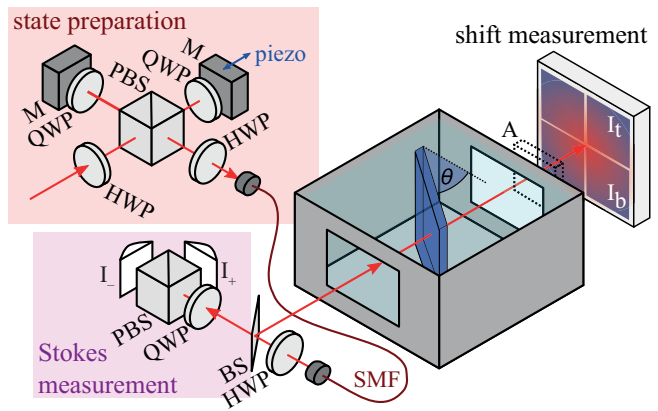}
 \caption{Experimental Setup:
Simultaneous measurement of the incident state of polarization and the
position of a light beam transmitted across a tilted polarizer.
State preparation:
The relative phase between horizontally and vertically polarized
components is modulated using a polarizing beam splitter (PBS), a piezo mirror
(M), quarter and half wave plates (QWP, HWP).
The beam is spatially filtered using a single-mode fibre (SMF).
Stokes measurement:
The transmitted port of a beam splitter (BS) is used to monitor the state of
polarization. We use the Stokes parameters $S_3$ to
distinguish left- and right-hand circular polarization.
Shift measurement:
The beam is propagated across our sample, a tank containing a glass
polarizer and an index-matching liquid, and its position is observed using a
quadrant detector.
An optional PBS (A) can be employed as an analyzer in front of
the detector.
The photo currents $I_{+}$, $I_{-}$, $I_t$, and
$I_b$ are amplified and digitally sampled for 1\,\second{} at 50\,\kilo\hertz.}
 \label{fig:setup}
\end{figure}

\begin{figure}
  \includegraphics[width=\figwidth]{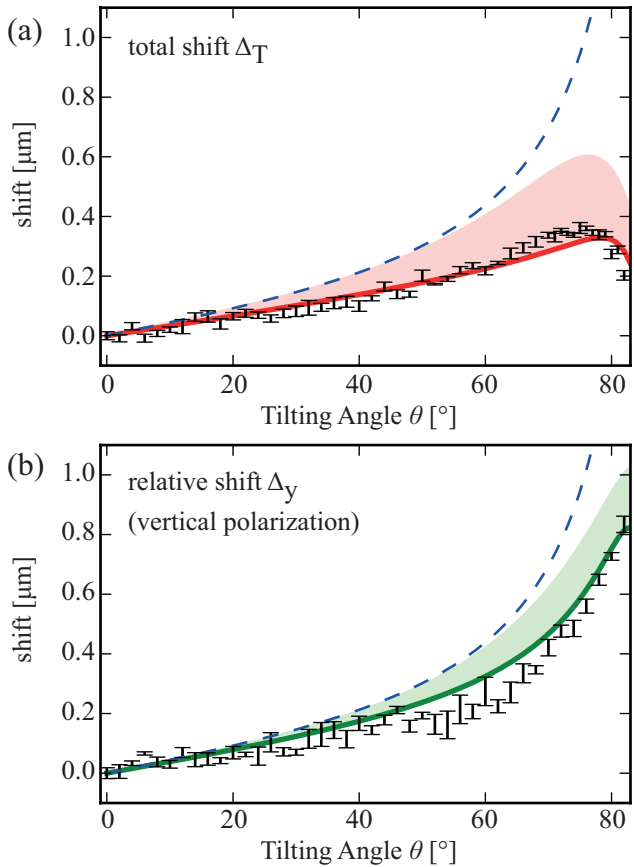}
 \caption{Polarization-dependent beam shifts occurring at a
tilted polarizer.
Measurement data and theoretical predictions are shown.
Both series of shift measurements were repeated five times. We report the
mean value and standard deviation of the mean.
The dashed blue line shows the theory for a perfect polarizer,
\CHANGE{while the shaded area indicates the result of our
phenomenological polarizer model for a reasonable range of parameters.
Details are given in the SM.}
(a)
Overall displacement $\Delta_T$ of the intensity barycentre after transmission
across the polarizer.
(b) Displacement $\Delta_y$ of the vertically polarized intensity component
solely.
The two measurements
$\Delta_T$ and $\Delta_y$ differ due to the imperfect nature of our
polarizer as discussed in the SM.
}
 \label{fig:shiftData}
\end{figure}

In our setup (\figref{setup}), a fundamental Gaussian light
beam ($\lambda = \SI{795}{\nano\metre}$)
is prepared with the state of polarization alternating between left-
and right-hand circular.
To avoid spatial jitter, the spatial mode is cleaned using
a single mode fibre and no active elements are used after the fibre.
We collimate the light beam using an aspheric lens (New Focus 5724-H-B)
aligned such that the beam waist is at the position of the
detector.

In order to simultaneously measure the beam position $\baryY{}$
and the incident state of polarization,
we employ a dielectric mirror (Layertec 103210) as a non-polarizing beam splitter
at an angle of incidence of $3\degree$.
The reflected and transmitted states of polarization
coincide
within experimental accuracy.
The beam centroid $\baryY{} = f\frac{I_t-I_b}{I_t-I_b}$ and Stokes parameter
$S_3 = \frac{I_+ - I_-}{I_+ + I_-}$ are calculated from the digitized photo currents.
We can measure the calibration factor $f$ in-situ by translating the quadrant detector using
a micrometre stage.

Since the signal is periodic with the modulation frequency $f\subrm{mod}
= 29\,\hertz$, 
we can filter technical noise in a post-processing stage.
To this end, the discrete Fourier transform is computed and only spectral components with
frequencies equal to $f\subrm{mod}$ and higher harmonics thereof are passed.

We identify
$S_3 = 0.99\pm0.01$ and $S_3=-0.99\pm0.01$
with the circular states of polarization $\sigma = +1$ and $\sigma = -1$
respectively. For both states, we calculate the mean of all corresponding
beam positions $\baryY{}\left(\sigma\right)$ and 
the helicity-dependent displacement
$\Delta^\mathrm{R} = \baryY{}\left({\sigma=+1}\right) -
\baryY{}\left(\sigma=-1\right)$.

An extensive characterization of statistical and systematic errors revealed
that the observed position of the light beam depends slightly on the state of polarization
even if no sample is present in the beam path.
The magnitude of this spurious beam shift is typically much smaller than the phenomenon, we intend to study.
In the Supplemental material, we discuss that this amounts to a small
offset \CHANGE{(60~nm)} on the raw data points $\Delta^\mathrm{R}(\theta)$, independent from
the action of the sample. Thus, the shift measurements $\Delta(\theta) = \Delta^\mathrm{R}(\theta) -
\Delta^\mathrm{R}(0\degree)$ reported here are corrected with respect to the raw
data.

We investigate beam shifts in two different
configurations.
First, we measure the displacement $\shiftT = 2\baryYET$
of the total transmitted energy density distribution $|\vec E_T|^2$ when switching
the incident state of polarization from $\sigma = +1$
to $\sigma = -1$  (\figref{shiftData}(a)).
Then, we employ an additional polarization analyzer in front of the detector
and observe the shift
$\shifty = 2\baryYE{y}$ of the energy density $|E_y|^2 = |\vec E_T\cdot\vech y|^2$ 
of the vertically polarized field component 
solely (\figref{shiftData}(b)).
These variants of the experiment coincide for polarizers with perfect extinction ratios
but can differ significantly for real-world polarizers with minor deficiencies.

The beam shift observed in the latter case increases
proportionally to the tangent of the tilting angle, exceeding one wavelength.
This characteristic ``$\tan\theta$''{-behaviour (and ``real-world-polarizer'' modification thereof)}
is unique to the geometric spin Hall effect of light. 
In both cases, the measurement agrees well with theoretical predictions
using our phenomenological model.

\begin{figure}
	\includegraphics[width=\figwidth]{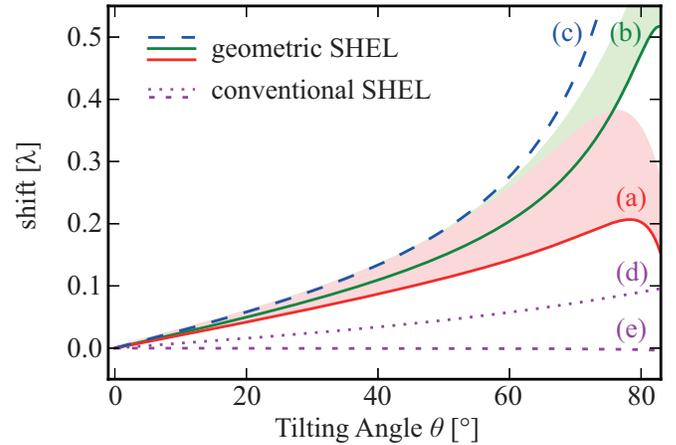}
	\caption{Theory for the conventional spin Hall effect of light compared to the geometric SHEL.
	A  light beam transmitted across an interface between two media can undergo
	a transverse displacement known as the Imbert-Fedorov effect or conventional SHEL.
	Here, we plot this displacement for a left-hand circularly polarized beam
	($\sigma = +1$) for two different cases and compare this with the geometric SHEL
	studied in this work.
	(a) and (b)
	Geometric SHEL $\frac12\Delta_T$ and $\frac12\Delta_y$ for the same configuration as in \figref{shiftData}.
	(c)
	Geometric SHEL as predicted for an ideal polarizing interface.
	(d)
	Conventional SHEL occurring at an air-glass interface (\nobreak{$n_1 = 1$}, \nobreak{$n_2=1.5$}).
	(e)
	Conventional SHEL expected for the entrance face of our submerged polarizer 
	($n_1 = n_L = 1.521$, $n_2 = n_G = 1.517$).
	}
	\label{fig:IF}
\end{figure}

To the best of our knowledge, this is the first direct measurement of this
intriguing phenomenon.
The geometric spin Hall effect of light should not be confused with the
conventional SHEL or Imbert-Fedorov shift.
The latter occurs at a physical interface
and, while such interfaces are
present in our experimental setup, they can only give rise to beam shifts
significantly smaller than the observed effect. In 
\figref{IF}, we compare our results to the Imbert-Fedorov shift,
which could occur at the polarizer substrate, for a set of
realistic parameters \cite{Bliokh2006,Liu2008}.
This illustrates that the beam shifts measured in this work constitute a novel
spin Hall effect of light, virtually independent from surface effects.

In conclusion,
we have demonstrated the 
geometric spin Hall effect of light experimentally by
propagating a circularly polarized laser beam across a suitable polarizing interface.
The centre of mass of the transmitted
light field was found to be displaced with respect to position of the incident beam
as predicted by the theory.
While a Gaussian light beam itself is invariant with respect to
rotation around its axis of propagation, the geometry induced
by the tilted polarizer, breaks this symmetry.
The resulting displacement can be interpreted as a spin-to-orbit coupling characteristic for spin Hall effects of light.

\section*{Acknowledgements}

The authors thank Christian Gabriel for fruitful discussions and for
his contribution in the initial stage of the experiment.

\clearpage
\section*{Supplemental Material}
\setcounter{equation}{0}
\renewcommand{\theequation}{S\arabic{equation}}
\setcounter{figure}{0}
\renewcommand{\thefigure}{S\arabic{figure}}

\newcommand{\eqShift}{eq.\ (2)}

\subsection{Phenomenological polarizer model and characterization of
  our real-world polarizer}

In this section, we describe a geometric polarizer model,
analogously to the work by Fainman and Shamir \cite{Fainman1984}, and
determine empirical parameters relevant for the actual polarizer used in
our experiment. Here, the interaction of an arbitrarily oriented polarizer with
a plane wave is discussed while we deal with real light beams in the
subsequent section on beam shifts.

A polarizer is an optical device that alters the state of polarization and
intensity of a plane wave without effecting its direction of propagation
$\vech\kappa=\vec k/k$.
The polarizer used within this work can be described by a real-valued unit vector $\vech P_a$
describing its absorbing axis.
Projecting $\vech P_a$ onto the transverse plane of the electric field yields
an effective absorbing axis
\begin{equation}
 \vech a(\vech\kappa) =
  \frac{\vech P_a - (\vech P_a\cdot\vech\kappa)\,\vech\kappa}
  {\sqrt{1 - (\vech P_a\cdot\vech\kappa)^2}}\text.
\end{equation}
Thus, the electric field transmitted across an idealized polarizer is
\begin{equation}
  \vect{E}_{T} 
  = \vect E_{I} - \vech a\left(\vech a\cdot \vect E_{I}\right)
  = \vech t\left(\vech t\cdot \vect E_{I}\right)\text,
\end{equation}
where $\vect E_I = \vect E_I(\vec k)$ is the amplitude of the plane wave
$\exp(i\,\vech\kappa\cdot\vec r)$, and
\begin{equation}
 \vech t(\vech\kappa) =
 \vech a(\vech\kappa) \times \vech\kappa
  = \frac{\vech P_a \times \vech\kappa}{|\vech P_a \times \vech\kappa|}
\end{equation}
is the effective transmitting axis.

\begin{figure}
  \includegraphics[width=\figwidth]{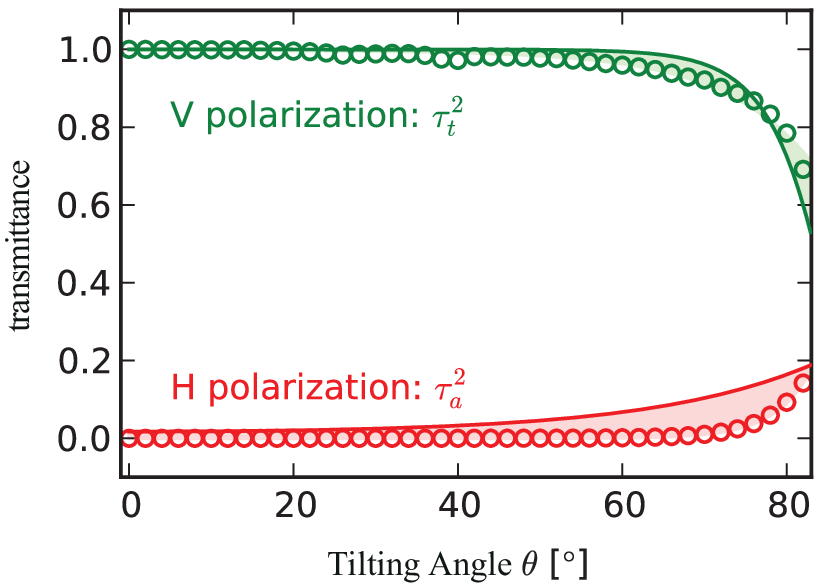}
 \caption{
\textbf{Transmittance of light beams across the polarizer 
as a function of the tilting angle $\theta$ and the incident state of 
polarization.} 
The polarizer is aligned such that at normal incidence ($\theta = 
0\degree$), vertical (V) polarization is transmitted and horizontal (H) 
polarization is blocked. 
We observe how the transmittance changes when the polarizer is rotated around 
the vertical axes and 
compare the experimental data (circles) to our phenomenological model
\eqref{eq:ET(EI)}
for different parameters $\tau_t$ and $\tau_a$.
The shaded regions indicate the range of parameters considered
within this work. One limit \eqref{eq:tau_ta} is depicted with solid lines
and shows the best fit with the observed beam shifts.
The other limit is the best fit to the measured transmittances.
}
 \label{fig:transData}
\end{figure}

However, real-world polarizers have finite extinction ratios and an
experimental characterization shows that the effectiveness of our polarizer
decreases when tilted (\figref{transData}).
Thus,
we have found it convenient to phenomenologically describe the transmitted light field as
\begin{equation}
 \label{eq:ET(EI)}
 \vect E_T = 
  \tau_t\,\vech t\left(\vech t\cdot\vect E_I\right) +
  \tau_a\,\vech a\left(\vech a\cdot\vect E_I\right)\text.
\end{equation}
Here, we employ two empirical parameters, 
$\tau_a(\theta)$ and $\tau_t(\theta)$,{}
depending on the angle $\theta$ between the propagation direction
$\vech\kappa$ and the unit vector perpendicular to the surface of the polarizer.
\CHANGE{
We have found the following set of parameters to be in very good agreement with the observed
transmission:
\begin{subequations}
\label{eq:tau_taT}
\begin{align}
 \tau_t(\theta) &= 1-0.31\,\exp(-5.6\,\cos\theta)\\
 \tau_a(\theta) &= 0.93\,\exp(-6.4\,\cos\theta)
\end{align}
\end{subequations}
This simple model is sufficient for our purpose, albeit it does not describe
the complex behaviour of the polarizer perfectly.}
A more detailed study of tilted polarizers can be found in Ref.~\cite{Korger2013_polarizer}.

In the following section, we will use this model to calculate beam shifts.
We have shown in the main text (Fig. 4), that this prediction matches qualitatively
to our observation.
Since the magnitude of these beam shifts depend critically on the 
parameters $\tau_t$ and $\tau_a$, a 
slightly different choice significantly increases the quantitative agreement:
\begin{subequations}
\label{eq:tau_ta}
\begin{align}
 \tau_t(\theta) &= 1-1.2\,\exp(-12\,\cos\theta)\\
 \tau_a(\theta) &= 0.51\,\exp(-1.3\,\cos\theta)
\end{align}
\end{subequations}
These parameters can be found by a combined fit of both, the observed transmission
and the beam shifts.
In figures 4, 5, and \ref{fig:transData}, the latter choice \eqref{eq:tau_ta} is depicted using
thick solid lines.
Shaded regions indicate a range of realistic parameters including both \eqref{eq:tau_taT}
and \eqref{eq:tau_ta} as limits.

\subsection{Calculation of beam shifts occurring at an oblique
polarizer according to our phenomenological model}

In this section, we adapt the theory for the geometric spin Hall effect of light
originally calculated for a different type of polarizer \cite{Korger2011} to our phenomenological 
model. 
To this end, we express the 
polarizer's absorbing axis
\begin{equation}
  \vech P_a = \vech x' = \cos\theta\,\vech x + \sin\theta\,\vech z
\end{equation}
in the global reference frame $\{\vech x, \vech y, \vech z\}$,
aligned with the direction $\vech z$ 
of beam propagation. 

The
incident beam $\vec E_I(\vec r)$ is circularly polarized and
well-collimated, i.e.\ it has a low divergence $\theta_0$,
with a Gaussian transverse intensity profile. 
This is expanded in a plane wave basis with amplitudes $\vect E_I(\vech\kappa)$ such that
the $\vech\kappa$-dependent projection \eqref{eq:ET(EI)} can be applied. 

As a consequence of our polarizer model,
the electric field $\vec E_T(\vec r)$ 
after transmission across such a polarizer, is 
a superposition of two orthogonally polarized field components,
$\vec E_{x}(\vec r)$ and $\vec E_{y}(\vec r)$.
Thus, the electric field energy density distribution (at the detection plane $z=0$)
\begin{equation}
\label{eq:ED=Ex+Ey}
\begin{split}
  |\vec E_T(x,y)|^2 &=
  \underbrace{|\vec E_T\cdot\vech x|^2}_{
  =|E_{x}|^2} +
  \underbrace{|\vec E_T\cdot\vech y|^2}_{
  =|E_{y}|^2} +
  \underbrace{|\vec E_T\cdot\vech z|^2}_{
  \approx 0}
\end{split}
\end{equation}
can be decomposed analogously.

Geometric SHEL manifests itself as a transverse displacement $\baryY{}$ 
of the transmitted light beam's barycentre.
The total electric energy density $|\vec E_T|^2$ and both of its non-vanishing components
undergo such a shift. 
Since this spatial displacement is independent from the $z$ coordinate,
we restrict the discussion to the detection plane at $z=0$.
It is convenient to write the total energy density's barycentre
\begin{align}
  \label{eq:baryY_wsum}
  \baryYET
  = w_{x}\,\baryYE{x}+w_{y}\,\baryYE{y}
\end{align}
as a weighted sum of the relative shifts occurring for the horizontally and
vertically polarized components respectively.
Here,
\begin{align}
  \baryY{u} = \frac{\IINT y\,u(x,y) \dif x\dif y}
    {\IINT u(x,y)\dif x\dif y}
\end{align}
denotes the centre of mass along the $\vech y$ direction
calculated with respect to a scalar distribution $u(x,y)$. The integration
spans the whole detection plane.

The weights
\newcommand{\weightDen}{\ensuremath{\tau_a^2(\theta)+ \tau_t^2(\theta)}}
\begin{align}
\begin{split}
 w_{x}(\theta) &= 
  \frac{\IINT E_{x}^{2}(x,y)\dif x\dif y}
  {\IINT |\vec E|^{2}(x,y)\dif x\dif y}
 \\
 &= \frac{\tau_a^2(\theta)}\weightDen + \OT2\quad\text{ and}
\end{split}\\
 w_{y}(\theta) &= \frac{\tau_t^2(\theta)}\weightDen + \OT2
\end{align}
introduced in \eqref{eq:baryY_wsum} depend on the
empirical parameters $\tau_{t}$ and $\tau_{a}$ \eqref{eq:tau_ta}.
Finally, we can calculate the relative shifts
\begin{align}
\label{eq:baryY_Ei}
\begin{split}
  \frac{\baryYE{i}}{\lambda} &=
  \frac{\IINT_{x,y} y\,\EDE{i} \dif x\dif y}
  {\IINT_{x,y} \,\EDE{i} \dif x\dif y}\\
  &= \sigma\,\frac{\tan\theta}{2\,\pi}\,f_{i}(\theta)+\OT2\text,
\end{split}
\end{align}
where the factors
\begin{subequations}
\begin{align}
 f_{x}(\theta) &= 1 - \frac{\tau_t(\theta)}{\tau_a(\theta)} < 0\quad\text{ and}\\
 f_{y}(\theta) &= 1 - \frac{\tau_a(\theta)}{\tau_t(\theta)} > 0
\end{align}
\end{subequations}
depend critically on the performance of the polarizer.

Note that, since the transmission coefficients are real and positive and $1 \geq \tau_t
> \tau_a$ (\figref{transData}), these relative shifts have opposite signs.
Consequently, the displacement of the total energy density
\begin{equation}
  \label{eq:baryY_final}
  \frac{\baryYET}\lambda =
  \sigma\,\frac{\tan\theta}{2\,\pi}\,
  \underbrace{\left(w_x\,f_{x} + w_y\,f_{y}\right)}_{<1}
\end{equation}
is smaller than the relative shift $\baryYE{y'}$ of the vertically polarized
component solely. For our realistic polarizer model, both shifts are smaller
than expected for the ideal case. This matches very well with our experimental
observation.

For a perfect polarizer with $\tau_a = 0$ and $\tau_t = 1$, \EQref{eq:baryY_final} reduces to
\eqShift\ 
since $w_x\,f_x = \frac{\tau_a^2}{\tau_t^2}\left(1 -\frac{\tau_t}{\tau_a}\right)\rightarrow0$
for $\tau_a\rightarrow0$.
This is exactly the same expression that was originally found for the polarizer model
by Fainman and Shamir \cite{Korger2011}.

\subsection{Calculation of the geometric spin Hall effect of light at a polarizing interface}

In this part, we derive detailed \eqShift\ for a paraxial 
fundamental Gaussian beam incident at an angle $\theta$ upon the 
surface of an \emph{absorbing polarizer} whose absorbing axis is directed along $x'$. The Cartesian coordinate systems attached to the polarizing interface and to the beam are  still defined as in Fig. 2A in the main text.
However, now the axis $x'$ is taken parallel to the absorbing axis of the polarizer, in order  to reproduce the actual experimental conditions. 

Consider the fundamental solution of the scalar paraxial wave equation 
\cite{Goodman} that we indicate with $\psi(\bm{r})$:
\begin{align}\label{SM100}
\psi(\bm{r}) = \left( \frac{k}{\pi L}  \right)^{1/2} \frac{i}{1+ i z/L}\exp \left( -\frac{1}{w_0^2} \frac{x^2 + y^2}{1+ i z/L}\right),
\end{align}
where the Rayleigh range $L$ of the beam can be expressed in terms of 
the beam waist $w_0$ as $L = k w_0^2/2$, with $\theta_0 = 2/(k w_0) \ll 
1$ denoting the angular spread of the beam. In the first-order 
approximation with respect to $\theta_0$, the electric vector field of 
the incident beam can be written as \cite{Haus1993}:
\begin{align}\label{SM110}
\bm{\Psi}^\text{in}(\bm{r}) &=  \vech{u}_\sigma \, \psi(\bm{r}) + \vech{z} \frac{i}{k} \vech{u}_\sigma \cdot \bm{\nabla} \psi(\bm{r}) \nonumber \\
&=  \left[
    \vech u_\sigma  -
    i \vech z \frac{\theta_0\,( x + i \sigma y)}{ \sqrt{2}\,w_0\,(1+i z/L)}
    \right]\psi(\vec r)\text,
\end{align}
where $\hat{\bm{u}}_\sigma = (\hat{\bm{x}}+i \sigma 
\hat{\bm{y}})/\sqrt{2}$ and an overall (irrelevant) multiplicative term 
has been omitted.
In Ref.\ \cite{Aiello2009d} it was demonstrated that the field transmitted by an arbitrarily oriented polarizer can be written as a perturbative expansion of the form 
\begin{equation}
\begin{split}\label{SM120}
\bm{\Psi}^\text{out}(\bm{r}) =&\; 
\widetilde{G}_{00}\bm{\Psi}^\text{in}(\bm{r})
\brOC
&- \frac{i}{k} \left[ 
\widetilde{G}_{10} \frac{\partial \bm{\Psi}^\text{in}}{\partial x}(\bm{r})  + \widetilde{G}_{01} \frac{\partial \bm{\Psi}^\text{in}}{\partial y}(\bm{r}) 
\right]
\brOC
&+ O(\theta_0^2),
\end{split}
\end{equation}
where  the $3 \times 3$ matrices $\widetilde{G}_{nm}$ are defined as
\begin{align}\label{SM130}
\widetilde{G}_{nm}= \left. \frac{k^{n+m}}{n! m!} \frac{\partial^{n+m}(\hat{\bm{n}}\hat{\bm{n}})}{\partial k_x^n \partial k_y^m }  \right|_{k_x=0, \; k_y=0}.
\end{align}
For our absorbing polarizer, the dyadic form 
\begin{equation}
  \vech n \vech n = \left(
  \begin{array}{ccc}
    n_x^2     & n_x\,n_y & n_x\,n_z\\
    n_y\,n_x & n_y^2     & n_y\,n_z\\
    n_z\,n_x & n_z\,n_y  & n_z^2
  \end{array}\right)
\end{equation}
is defined in terms of the effective-transmission unit vector
$\vech{{n}} = \vech{{a}} \times \vech{{\kappa}}
=n_x\vech x + n_y\vech y + n_z\vech z$,
where 
\begin{align}\label{SM140}
\hat{\bm{a}} = \frac{\hat{\bm{x}}' - \hat{\bm{\kappa}} ( \hat{\bm{\kappa}}  \cdot \hat{\bm{x}}' )}{\sqrt{1- |\hat{\bm{\kappa}}  \cdot \hat{\bm{x}}'|^2}} ,
\end{align}
and $\hat{\bm{\kappa}} = \bm{k}/k$.
A straightforward calculation furnishes
\begin{subequations}
\label{eq:G}
\begin{equation}
\label{eq:Ga}
\widetilde{G}_{00} = \left(
\begin{array}{ccc}
 0 & 0 & 0 \\
 0 & 1 & 0 \\
 0 & 0 & 0
\end{array}
\right),
\end{equation}
\begin{equation}
\label{eq:Gb}
\widetilde{G}_{01} = \left(
\begin{array}{ccc}
 0 & \tan \theta & 0 \\
 \tan \theta & 0 & -1 \\
 0 & -1 & 0
\end{array}
\right)\text{,}
\end{equation}
\begin{equation}
\label{eq:Gc}
\widetilde{G}_{10} = \left(
\begin{array}{ccc}
 0 & 0 & 0 \\
 0 & 0 & 0 \\
 0 & 0 & 0
\end{array}
\right)\text.
\end{equation}
\end{subequations}

Substitution of \EQsref{eq:G} and \eqref{SM110} into \EQref{SM120} yields to the following first-order expression for the transmitted field: 
\begin{equation}\begin{split}
\bm{\Psi}^\text{out}(\bm{r}) =
\frac{\psi(\bm{r})}{\sqrt{2}} \bigg[ 
&- \hat{\bm{x}} \, \theta_0 \frac{y}{w_0} \frac{\sigma \tan \theta}{1 + i z/L}
\brOC
&+ \hat{\bm{y}}\,i \sigma  \left( 1 + \theta_0 \frac{y}{w_0} \frac{\sigma \tan \theta}{1 + i z/L}\right) \\
&+ \hat{\bm{z}} \, \theta_0 \frac{y}{w_0} \frac{\sigma }{1 + i z/L} 
\bigg]
+ O(\theta_0^2).
\end{split}\end{equation}
The electric field energy density, the physical quantity actually measured by a standard optical detector, is proportional to $|\bm{\Psi}^\text{out}(\bm{r})|^2$ which can be calculated from Eq. \eqref{SM140} as:
\begin{align}\label{SM150}
|\bm{\Psi}^\text{out}(\bm{r})|^2 = |\psi(\bm{r})|^2 \left( \frac{1}{2}  +  \theta_0 \frac{y}{w_0} \frac{\sigma \tan \theta}{1 + i z/L}
 \right) + O(\theta_0^2).
\end{align}
Finally, the sought shift  is calculated as the first moment of the 
electric field energy density distribution, namely
\begin{align}\label{SM160}
\langle y \rangle = & \; \frac{\displaystyle{\int \! \! \! \! \int y \, |\bm{\Psi}^\text{out}(\bm{r})|^2 \text{d} x \text{d} y}}{\displaystyle{\int \! \! \! \! \int |\bm{\Psi}^\text{out}(\bm{r})|^2 \text{d} x \text{d} y}} \nonumber \\
  = & \; \frac{\sigma}{k} \tan \theta,
\end{align}
where the integration is evaluated over all the $xy$ plane at $z = 0$. 
Equation \eqref{SM160} reproduces the result given by \eqShift\ in 
the main text.

\subsection{Discussion of spurious beam shifts}

In our experiment, we prepare left and right circularly polarized 
Gaussian light beams and intend to study how a sample effects 
the position of the transmitted beam in both cases.
The spatial modes prepared in each case are almost, 
yet not exactly identical. Despite being filtered using 
two metres of single-mode fibre, there can be a tiny offset between the
centre of mass of the left versus the right circularly polarized beam
prepared in this manner. While this effect is too small to be imaged
on a CCD camera,  it can be revealed in a precision measurement as employed
in this work.

Theoretically, we can account for such spurious beam shifts by 
substituting the ideal Gaussian profile $\vect E_I(\vech\kappa)$ with a generic
paraxial beam such as a power series or superposition of Hermite-Gaussian
laser modes.
This changes the result \eqref{eq:baryY_Ei} to
\begin{equation}
  \frac{\baryYE{i}}{\lambda} =
  \sigma\,\frac{\tan\theta}{2\,\pi}\,f_{i}(\theta)+Y_i(\sigma)+\OT2\text,
\end{equation}
where $Y_i(\sigma)$ is an offset independent of the tilting angle $\theta$,
which can also be understood intuitively: The deviation between two almost
identical beam profiles amounts to either a spatial or an angular beam shift.
First, since the sample is spatially homogeneous, a spatial displacement of the whole beam,
does not change the interaction at all.
And second, while the interaction with the sample generally depends on the angle of incidence, a slightly different
direction of propagation,
here on the order of $10^{-6}~\rad$, is practically negligible. 

In our experiment, we measure the helicity-dependent beam shift
\begin{equation}
\begin{split}
  \Delta^\mathrm{R}
  &= \baryY{}\left({\sigma=+1}\right) -
  \baryY{}\left(\sigma=-1\right)
  \\&= \underbrace{2\frac{\tan\theta}{2\,\pi}\,f_{i}(\theta)}_{
  \Delta(\Theta)}
  +\underbrace{Y_i(\sigma=+1)-Y_i(\sigma=-1)}_{
  \Delta^\mathrm{R}(\theta=0\degree)}
  \text,
\end{split}
\end{equation}
which contains a spurious offset $\Delta^\mathrm{R}(\theta=0\degree)\approx 60~\nano\metre$.
The data shown in Fig. 4 is corrected for this offset.

Such offset can be roughly quantified in a simple manner.
In our experimental conditions, the used single-mode fiber  does not support exactly a single mode only, namely the fundamental $\text{TEM}_{00}$ mode, but also first-order modes such as $\text{TEM}_{10}$ and $\text{TEM}_{01}$.
Ideally, a conventional single-mode optical fiber possesses a simple circular cylindrical geometry in absence of either intentional or accidental deformations of the fiber structure. Any deviations from the ideal configuration causes proliferation of, and coupling between, normal modes of the fiber.  Typically, a small coupling may occur because of several reasons, including intrinsic or induced cross-sectional ellipticity, bending, twist, etc., of the fiber cross section
\cite{Hung83}.
Now, it is well known that the coupling between fundamental and first-order modes causes a shift of the centroid of the intensity distribution of the guided light inside the fiber
\cite{Hsu2004}.
This can be seen by writing the electric field distribution $E(x,y)$ (with a fixed uniform polarization) of the guided light, as a superposition of the $\text{TEM}_{00}$ and the $\text{TEM}_{01}$ mode
\begin{align}
E(x,y) \approx \text{TEM}_{00}(x,y) + \eta \, \text{TEM}_{01}(x,y),
\end{align}
where $0 <\eta \leq 1$ is the dimensionless coupling parameter.
The spatial intensity distribution of such a field configuration is displaced along the $y$-axis by $\Delta y$, where
\begin{align}\label{estimate}
\Delta y = & \; \frac{\displaystyle{\iint y \left| E(x,y) \right|^2 \, \text{d}x \text{d}y}}{\displaystyle{\iint  \left| E(x,y) \right|^2 \, \text{d}x \text{d}y}} \nonumber \\
= & \; w_0 \, \frac{\eta}{1+\eta^2} \nonumber \\
\simeq & \; w_0 \, \eta,
\end{align}
where the last, approximate equality holds in the case $\eta \ll 1$ and $2 w_0$ is the diameter of the fundamental $\text{TEM}_{00}(x,y)$ mode of the fiber.
In our experimental conditions, we had $w_0 = 2.5 \, \mu\text{m}$  and 
we measured $\Delta y \simeq 60 \, \text{nm}$.   By inserting these 
values into Eq. \eqref{estimate} we obtain $\eta \simeq \Delta y / w_0 
\simeq 0.024$. This number tells us that it is enough to have an amplitude ratio of about $2\%$ between the first-order $\text{TEM}_{01}$ and the  fundamental $\text{TEM}_{00}$ mode, to observe a transverse shift of $60 \, \text{nm}$.

For the present rough estimation, the use of a single scalar coupling coefficient $\eta$ is appropriate. However, if one would adhere more to the physical reality of the  phenomenon, one would rather use a tensor coupling coefficient $\eta_{ij}$ in order to account for the polarization-dependence of the coupling.

\end{document}